\pdfoutput=1
\documentclass[pra,twocolumn]{revtex4}
\usepackage{epsfig}
\usepackage{graphicx}
\usepackage{amsmath}
\usepackage{natbib}
\usepackage{color}

\newcommand{\n}{\nonumber}
\newcommand{\bn}{\begin{eqnarray}}
\newcommand{\en}{\end{eqnarray}}
\newcommand{\eml}{\end{multline}}
\newcommand{\bml}{\begin{multline}}
\newcommand{\h}{\hspace}

\begin{document}

\title {Realizing the Harper model with Ultracold Atoms in a Ring Lattice}

 \author{Kunal K. Das$^{1,2}$ and Jacob Christ$^{1}$ }
 \affiliation{
$^{1}$Department of Physical Sciences, Kutztown University of Pennsylvania, Kutztown, Pennsylvania 19530, USA\\
$^{2}$Department of Physics and Astronomy, Stony Brook University, Stony Brook, New York 11794-3800, USA}

\date{\today }
\begin{abstract} We demonstrate that all of the salient features of the Harper-Hofstadter model can be implemented with ultracold atoms trapped in a bichromatic ring-shaped lattice.  Using realistic sinusoidal lattice potentials rather than assume the idealized tight-binding picture, we determine the optimal conditions necessary to realize the critical point where the spectrum becomes fractal, and identify the nature and cause of the departures from the discrete model predictions.  We also show that even with a commensurate ring with a few lattice sites, the Aubry-Andr\'e localization transition can be realized.  Localized states that behave like edge states with energies that reside in the band gaps can be generated by introducing a surprisingly small local perturbation within the ring. Spectrum oscillation arising from complex coupling can be implemented by uniform rotation of the ring, but with certain significant differences that are explained. \end{abstract}
%

\maketitle

\section{Introduction}

Consideration of electrons in a two-dimensional (2D) lattice subject to a magnetic field led Harper \cite{Harper} to his eponymous model which has since been the subject of a vast number of studies, that continue unabated till the present day \cite{Kim-Hofstadter-expt-electrons, Geim-Hofstadter-expt-electrons}. It has been an essential part of the physics of the quantum Hall effect \cite{Stone-QHE, Prange-QHE}  and of recognizing the significance of topology in quantum physics \cite{TKNN, Thouless-book,Topology-review}, which has been transformational for our understanding. In recent years there has been tremendous interest in replicating associated phenomena in designer systems of ultracold atoms in the context of synthetic gauge fields and topological structures for neutral atoms \cite{RMP-Dalibard, Review-Goldman,Spielman-review}.

The Harper model can be famously reduced to an effective 1D Hamiltonian with nearest neighbor coupling and a cosine modulation of the onsite energies,
\bn J_1[e^{i\vartheta}\psi_{n+1} + e^{-i\vartheta}\psi_{n-1}] + J_2\psi_n \cos(2\pi\alpha n+\theta)=E\psi_{n}.\label{discrete-Hamiltonian}
\en
Here $J_1$ and $J_2$ represent the strengths of the coupling and the  modulation, while the phases $\theta$ and $\vartheta$ can be related to the wavenumbers in the 2D system. When the parameter $\alpha$ is irrational, the lattice index $n$ has infinite range \cite{Hofstadter-PRB}. But, when  it is rational, $\alpha=p/q$ with integer $p,q$, the Hamiltonian is of period $q$, in which case, the system can be mapped to a 1D ring-shaped lattice.

Although this mapping with dimensional reduction has been an intrinsic part of the Harper model, it is yet to be utilized in a literal sense in experiments, which have remained anchored in the 2D configuration of its genesis. That applies to even recent studies with ultracold atoms in optical lattices \cite{Jaksch-Zoller,Ketterle-butterfly, Bloch-butterfly}. However, precisely in this last realm the capability has now emerged that would enable experimental realization of this seminal model in its reduced dimensional representation: Numerous experiments have already been done with ultracold atoms in ring-shaped traps \cite{Boshier-painted-potential,Phillips-2007,ramanathan,Phillips_Campbell_hysteresis,Hadzibabic-spinor,Stamper-Kurn-2015}, and periodic lattice structure along the azimuth have also been demonstrated \cite{Franke-Arnold:07,Franke-Arnold:08,Zielinger-OAM,Kwek-ring-lattice,Abraham-LG}.  The ring lattice actually provides a simpler and possibly better alternative to examine this model. The 2D model is intrinsically finite with edges, and additional potentials required for confinement can introduce inhomogeneity not present in the classic Harper model. In contrast, a ring represents an infinite 2D system exactly without the complications of edges (although they can be easily introduced if desired), without any extra confinement required along the direction of relevant dynamics.

It is to motivate and anticipate such experiments with ultracold atoms, this study has been undertaken. The discrete Hamiltonian above is an idealization in the tight binding limit, and therefore well-established results that are derived from it will certainly be modified and distorted when real potentials are used. The goal of this paper is thus threefold: (1) to establish that all the salient features of the Harper model can indeed be implemented with ultracold atoms on a continuum ring-shaped lattice with realistic potentials that do not assume a tight-binding model, (2) to determine the optimal conditions under which the discrete model results can be reproduced, and (3) to identify the differences from the idealized discrete model, that emerge and linger in the continuum model.

\section{Continuum Model and Spectrum}

We will translate the model represented in Eq.~(\ref{discrete-Hamiltonian}) to a ring-shaped lattice described by a continuum Hamiltonian, $H$ with a bichromatic potential involving two independent sinusoidal modulations, given by
\bn H&=&H_0+V_H\cos\left( \frac{2\pi\alpha x}{a}+\theta\right)\n\\
H_0&=&-\frac{\hbar^2}{2m}\frac{d^2}{d x^2}+V_L \sin^2\left(\frac{\pi x}{a}\right)\label{continuum-Hamiltonian}\en
Here, $x$ measures position along the azimuth of the ring that will contain all relevant dynamics examined here, with the assumption of tight confinement along the other two degrees of freedom.  The phase $\theta$ will be set to zero until Sec. VI where we will describe some of its influence, and we will introduce the counterpart of the phase $\vartheta$ of the coupling coefficient in Sec.~VII.

Such a bichromatic model was utilized in an experiment demonstrating a localization transition for a Bose-Einstein condensate (BEC) in a harmonically confined lattice with incommensurate periods  \cite{Roati-localization}, and was recently studied to examine mobility edges and localization properties in an open incommensurate lattice \cite{Das-Sarma-mobility}. In contrast, due to the ring geometry, the two potentials here will be chosen to be commensurate. While there is flexibility depending on the physical configuration, for the sake of having a concrete picture, we will assume a torus of cylindrical cross-section with the lattice potentials along its major axis. Unless otherwise specified, the lattice constant $a$ and  $\epsilon_0=2 E_R/\pi^2$ will set the length and energy units, with $E_R=\hbar^2\pi^2/(2ma^2)$ being the recoil energy. We will neglect any non-linearity due to atom-atom interaction, assuming low density or scattering length manipulation by Feshbach resonance \cite{RMP-Feshbach}.

The amplitude of the potential that creates the base lattice structure is denoted $V_L$, the separation between the minima, $a$, corresponding to the lattice constant in the discrete model. The parameter $V_H$ is the coefficient of the Harper modulating potential and can be identified with $J_2$ in Eq.~(\ref{discrete-Hamiltonian}). However, to find the counterpart for $J_1$, we need to compute the overlap integral of localized states in adjacent sites.  For this purpose we will neglect the modulating potential since it varies with $\alpha$, but more importantly, as we will see, it will be relatively much smaller for cases of interest. Thus, we define the onsite energy and the nearest neighbor overlap integral as
\bn {\cal E} =\langle \phi_n| H_0|\phi_n\rangle \h{1cm}
\Delta=\langle \phi_n| H_0|\phi_{n+1}\rangle \label{integrals}\en
where $\phi_n$ denotes the state localized at lattice $n$, which in our calculations will correspond to the Wannier state for the lowest band for the unmodulated Hamiltonian $H_0$.

Comparison of the spectrum of the continuum Hamiltonian with that of the discrete Hamiltonian therefore entails the follow transformation
\bn H\rightarrow (H-{\cal E})/\Delta\h{1cm} E_i\rightarrow (E_i-{\cal E})/\Delta\en
for the Hamiltonian and its eigenenergies. In our plots, we present the eigenenergies as transformed above.

The most well known feature of the Harper model is a fractal spectrum known as the Hofstadter butterfly \cite{Hofstadter-PRB} which corresponds to a special case of the Harper Hamiltonian, when the ratio $\lambda_d=J_2/J_1=2$ in the discrete (subscript `$d$') model.   With our definition above, the equivalent for that ratio in the continuum (subscript `$c$') model is $\lambda_c=V_H/\Delta$, and so we computed the spectrum of the Hamiltonian in (\ref{continuum-Hamiltonian}) for the special value $\lambda_c=2$.

\begin{figure}\includegraphics[width=\columnwidth]{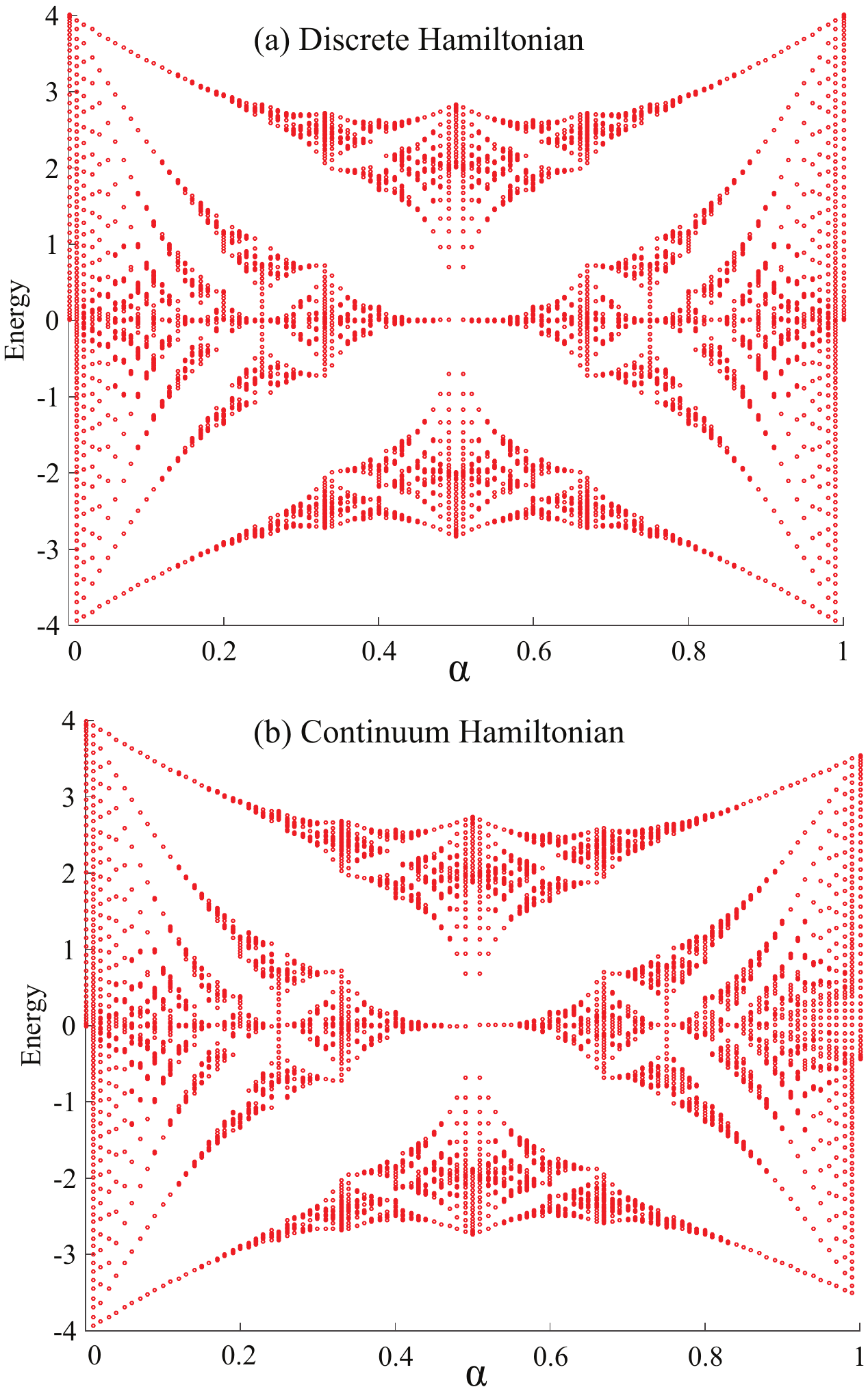}
\caption{The Hofstadter butterfly spectrum is found to be almost identical as generated with (a) the usual discrete Hamiltonian in Eq.~(\ref{discrete-Hamiltonian}), and (b) with the continuum Hamiltonian using the sinusoidal bichromatic potential in Eq.~(\ref{continuum-Hamiltonian}). The parameter $\alpha$ is dimensionless and the energy is scaled in units of the nearest neighbor couplings, $J_1$ in (a) and by $\Delta$ in (b). }\label{Figure-1-spectrum}
\end{figure}

With an optimal choice of lattice parameters, discussed below, the continuum Hamiltonian Eq.~(\ref{continuum-Hamiltonian}) on a ring-shaped lattice can reproduce the Hofstadter butterfly spectrum, almost indistinguishable from that generated with the discrete Hamiltonian Eq.~(\ref{discrete-Hamiltonian}), as we show in Fig.~\ref{Figure-1-spectrum}. Here, as well in the rest of the paper unless otherwise mentioned, the number of lattices sites, or potential minima of the primary lattice used is $N=100$, so the spectrum has a domain of $\alpha=n/N, n\in [1,100]$.

\section{Parametric Trade-off}

\begin{figure}[t]
\includegraphics[width=\columnwidth]{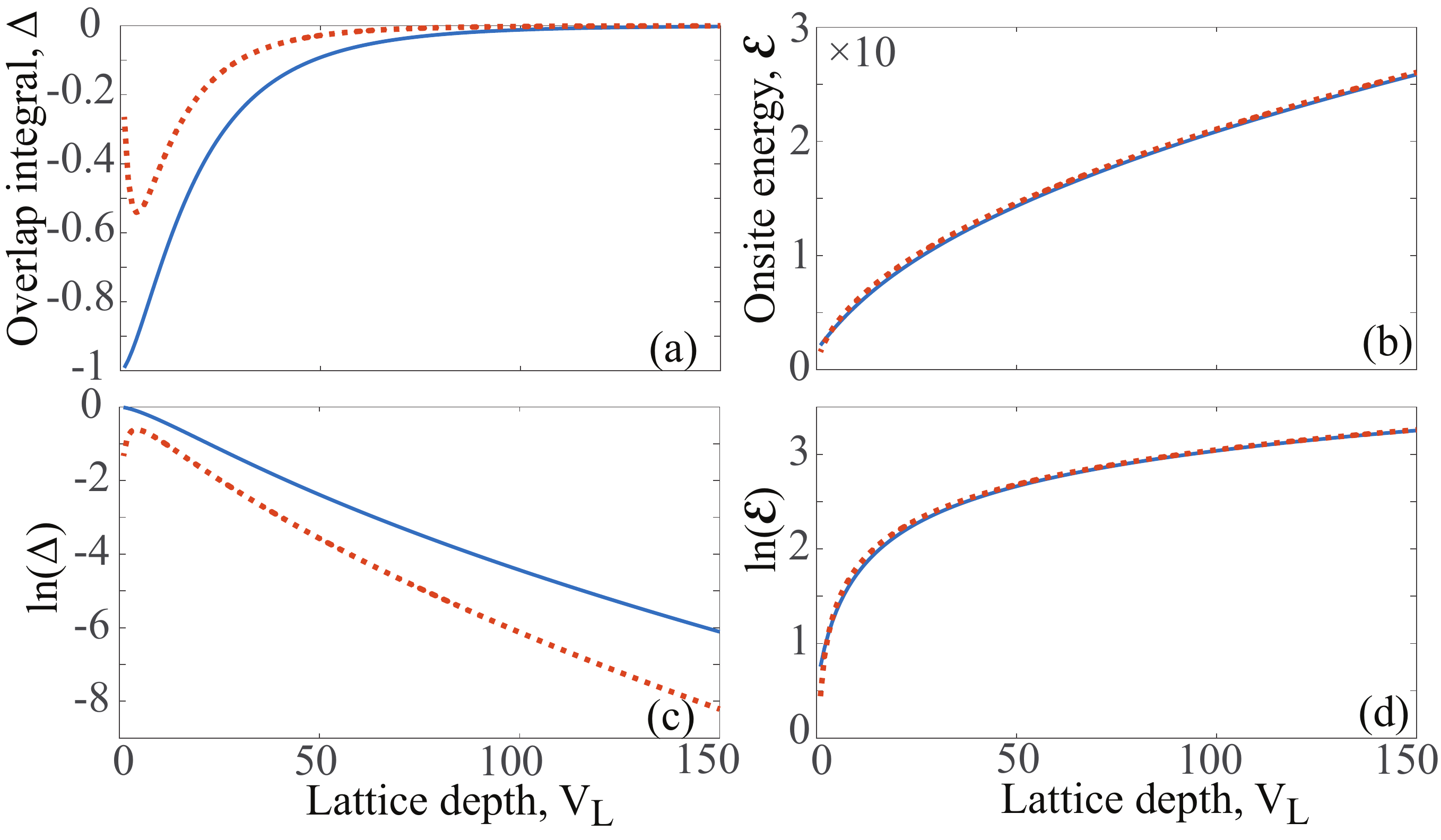}
\caption{(a) The overlap integral and (b) the onsite energy plotted as a function of the amplitude of the primary lattice potential $V_L$ computed using Eq.~(\ref{integrals}).  The solid blue line uses the lowest Wannier states for $\phi_n$ at neighboring sites and the dashed red line represents analytical computation using the ground state  $\phi_n^{HO}$ of the harmonic oscillator approximation of a well-bottom of the primary lattice. Semi-log plots of the same are shown in figures (c) and (d) respectively.}\label{Figure-2-energy-coupling}
\end{figure}

Given the greater degrees of freedom available, obtaining a well-defined fractal spectrum with the continuum Hamiltonian depends significantly on the lattice parameters.  For this purpose, we computed the onsite energy ${\cal E}$ and the hopping energy $\Delta$  as a function of the depth $V_L$ of the primary lattice, using the lowest band Wannier functions. For comparison, we also computed the same by approximating the well-bottom of the primary lattice by a harmonic oscillator of frequency $\omega=(\pi/a)\sqrt{2V_L}$, and using its ground state $\phi_n\rightarrow \phi_n^{HO}$ in Eq.~(\ref{integrals}) to analytically evaluate counterparts ${\cal E}^{HO}$ and $\Delta^{HO}$. We plot them all in Fig.~\ref{Figure-2-energy-coupling}. It is clear that harmonic oscillator approximation works well for the onsite energy $\cal E$, but is inaccurate for the more relevant overlap integral, $\Delta$. On the semi-log plots it is evident that for $\Delta$, the difference remains significant for all values of the lattice depth. We therefore only use the Wannier functions in our calculations.

As the lattice gets deeper, the system approaches the discrete limit, as the tight-binding picture gets increasingly precise, so it may seem that deeper the base lattice, the better it is. However, there is a trade-off, because the nearest-neighbor coupling measured by $\Delta$ decreases exponentially at higher lattice depths as evident from the Fig.~\ref{Figure-2-energy-coupling}(c). Considering the criterion for the Hofstadter spectrum, in the regime of interest, $V_H\sim\Delta$, which means the modulating potential $V_H$ has to decrease in sync with $\Delta$. At high lattice depths, this would result in a huge difference in magnitudes of $V_L$ and $V_H$ which could be challenging in experiments, particularly if the mean magnitude of the latter becomes comparable to the fluctuations of the former.

\begin{figure}[t]
\includegraphics[width=\columnwidth]{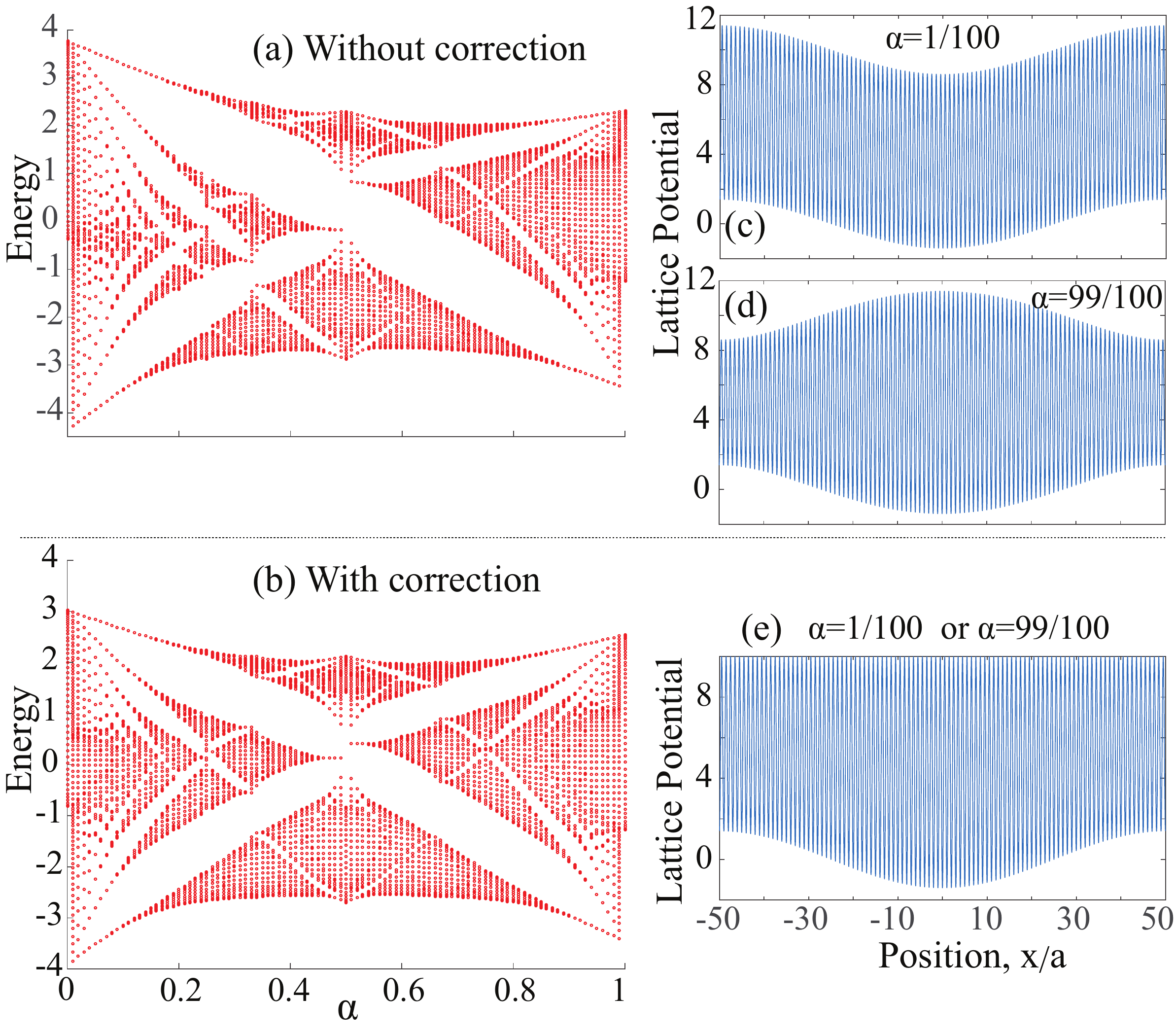}
\caption{(a) Distorted butterfly when the primary lattice depth is low $V_L=10$. (b) When the same is plotted with the correction potential $V_{corr}$ in Eq.~(\ref{correction}) added to the Hamiltonian $H_0$, the bilateral symmetry is partially restored. The uncorrected net potential for (c) $\alpha=1/N$ and (d) $\alpha=(N-1)/N$, (with $N=100$ lattice sites), are clearly different, but they acquire the same form shown in (e) when $V_{corr}$ is added. }\label{Figure-3-correction}
\end{figure}

In Fig.~\ref{Figure-1-spectrum} we used $V_L=100$ which yielded $\Delta=0.012$ and $V_H=2\Delta=0.024$. This is already a difference of $\log(V_L/V_H)\simeq 3.6$ orders of magnitude. There, we pushed the limits to demonstrate the reproducibility of the discrete model, and that may not be always be a priority. Even reducing the primary lattice depth by a factor of 2, to $V_L=50$ leads to critical $V_H=0.092$ and $\log(V_L/V_H)\simeq 2.4$, an one-third reduction in the order of magnitude gap. Certainly this creates more deviation from discrete spectrum, but not markedly. But, if $V_L$ gets too small, the tight-binding approximation is no longer a good one and the spectrum deviates significantly from that of the discrete model spectrum, a matter we address in the next section. So, some intermediate value of the primary lattice will have to be chosen, which provides the desirable balance between these opposing factors.

\begin{figure*}\includegraphics[width=\textwidth]{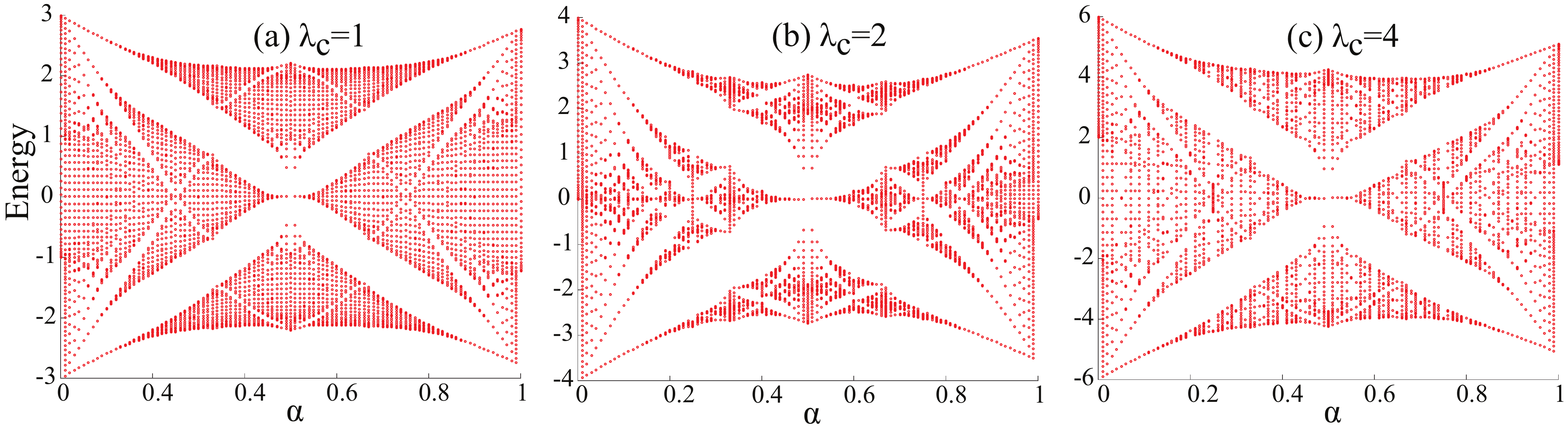}
\caption{The Hofstadter butterfly spectrum generated with the continuum Hamiltonian in Eq.~(\ref{continuum-Hamiltonian}). As predicted for the discrete model, the fractal structure is most pronounced for the critical ratio (b) $\lambda_c=V_H/\Delta=2$ in the center panel and deteriorates for values (a) lower and (c) higher.}\label{Figure-4-criticality}
\end{figure*}

\section{Distortions in Shallow Wells}

It is evident in Fig.~\ref{Figure-1-spectrum} that even though the spectra match quite well, there is a lingering bilateral asymmetry in the continuum case, particularly prominent near the right edge towards $\alpha=1$. This gets more pronounced for lower values of $V_L$ when the primary lattice gets shallower as evident in Fig.~\ref{Figure-3-correction} where we use $V_L=10$.  That is because the discrete Hamiltonian contains a symmetry absent in the continuum Hamiltonian: The Hamiltonian in Eq.~(\ref{discrete-Hamiltonian}) is unchanged by $\alpha=n/N\rightarrow (N-n)/N$, but that is not so in the continuum,
\bn \cos\left( \frac{2\pi x}{a}\frac{(N-n)}{N}\right)=\cos\left( \frac{2\pi x}{a} -\frac{2\pi x}{a}\frac{n}{N}\right).\en
The continuous dependence on the position $x$ modulates the overall potential differently for $\alpha=n/N$ and $\alpha=(N-n)/N$ as can be clearly seen on comparing Figs.~\ref{Figure-3-correction}(c) and (d) where $\alpha=1/100$ and $\alpha=99/100$ respectively. Note that for both cases, the overall modulation of the bottom edge are the same, and that is precisely what would be manifest as modulation of the onsite energies in the discrete model. But the difference in their upper edges indicates that the lattice amplitude has different behavior among the two. There can be variation of the lattice depth, and hence the nearest-neighbor coupling, across the lattice, evident particularly in Fig.~\ref{Figure-3-correction} (d), a feature clearly absent in the discrete model, where only the onsite energy is modulated.

Notably, the spectrum in Fig.~\ref{Figure-3-correction}(a) shows that the features on the right side of the spectrum are more degraded compared to the left side.  The reason why becomes apparent on comparing Figs.~\ref{Figure-3-correction}(c) and (d). For lower values of $\alpha$ there is little modulation of the lattice depth since the top and the bottom edges rise and fall in sync, and the nearest-neighbor coupling remains uniform across the lattice. On the other hand the lattice depth clearly varies significantly for the higher values of $\alpha$. The underlying reason resides in the fact that in the Harper modulation $\cos(2\pi\alpha x/a)$,  for low values of $\alpha$ its period $\alpha^{-1}a$ is much longer than the period $2a$ of the primary lattice, whereas for higher values of $\alpha$ the two periods become comparable, leading to a beating effect which modulates the net lattice depth, and hence the coupling strength, across the lattice.  This suggests that using low values of $\alpha$ would be preferable for studies that are not $\alpha$-specific.

When $V_L\gg V_H$ the distortion is minimal, but is pronounced when $V_L$ is smaller due to the greater relative impact of the cosine term, for instance, in Fig.~\ref{Figure-3-correction}, $V_L=10$ and $V_H=0.70$ and $\log(V_L/V_H)\simeq 0.85$. To underscore the points made above, we partially compensate for the asymmetry  with a correction term added to the potential:
\bn V_{corr}=-V_H\cos\left( \frac{2\pi\alpha x}{a}\right)\sin^2\left(\frac{\pi x}{a}\right)\label{correction}\en
This amounts to subtracting the Harper modulation, but further modulated by the periodicity of the primary lattice. It levels out the upper edge of the net lattice potential as shown in Figs.~\ref{Figure-3-correction} (e) and substantially reduces the asymmetry in the vertical spread of the eigenvalues between the left and the right sides of the butterfly.  On the other hand, this extra potential also reduces the clarity and resolution of the fractal pattern on the left side by causing lattice depth modulation there as well. As such, such a term may not offer any practical advantage. But, it serves to demonstrate how some of the unwanted features could be selectively neutralized, as well as illustrating some of the limitations and differences and their causes, that mark the continuum implementation.

\section{Criticality and Localization}

The fractal pattern of the Hofstadter butterfly is specific to the critical case of the Harper Hamiltonian, when the ratio $\lambda_d=2$ and deteriorates away from the value. Utilizing the continuum counterpart of the ratio $\lambda_c=V_H/\Delta$, we find that this critical behavior can be faithfully replicated on the bichromatic ring lattice.  This is shown in Fig.~\ref{Figure-4-criticality}  where we plot the spectra for the cases  $\lambda_c=1,2$ and $4$, and we can confirm that the most well-defined fractal pattern results when the $\lambda_c=2$ and the pattern gets smudged for values both lesser and greater.

In the Harper Hamiltonian, the value  $\lambda_d=2$ has a significance beyond the nature of the spectrum. It has been proven by Aubry and Andr\'e \cite{Aubry-Andre} that in the infinite lattice limit, when $\alpha$ is irrational, this value marks a localization transition, all eigenstates being localized for $\lambda_d>2$ and extended for $\lambda_d<2$. But, rigorous proof is lacking for commensurate finite lattice periods. Here we show that the transition does exist even for a ring-lattice of few sites with intrinsically commensurate periods. Infinite range lattice corresponds to irrational values of $\alpha$ and strictly irrational values are unfeasible in practice. A common choice for a rational alternative has been to pick a ratio of a pair of adjacent Fibonacci numbers $\alpha=F_n/F_{n+1}$ because the limit of the sequence as $n\rightarrow \infty$ is a well-known irrational number, $(\sqrt{5}-1)/2$, the inverse of the golden mean.

\begin{figure}[t]
\includegraphics[width=\columnwidth]{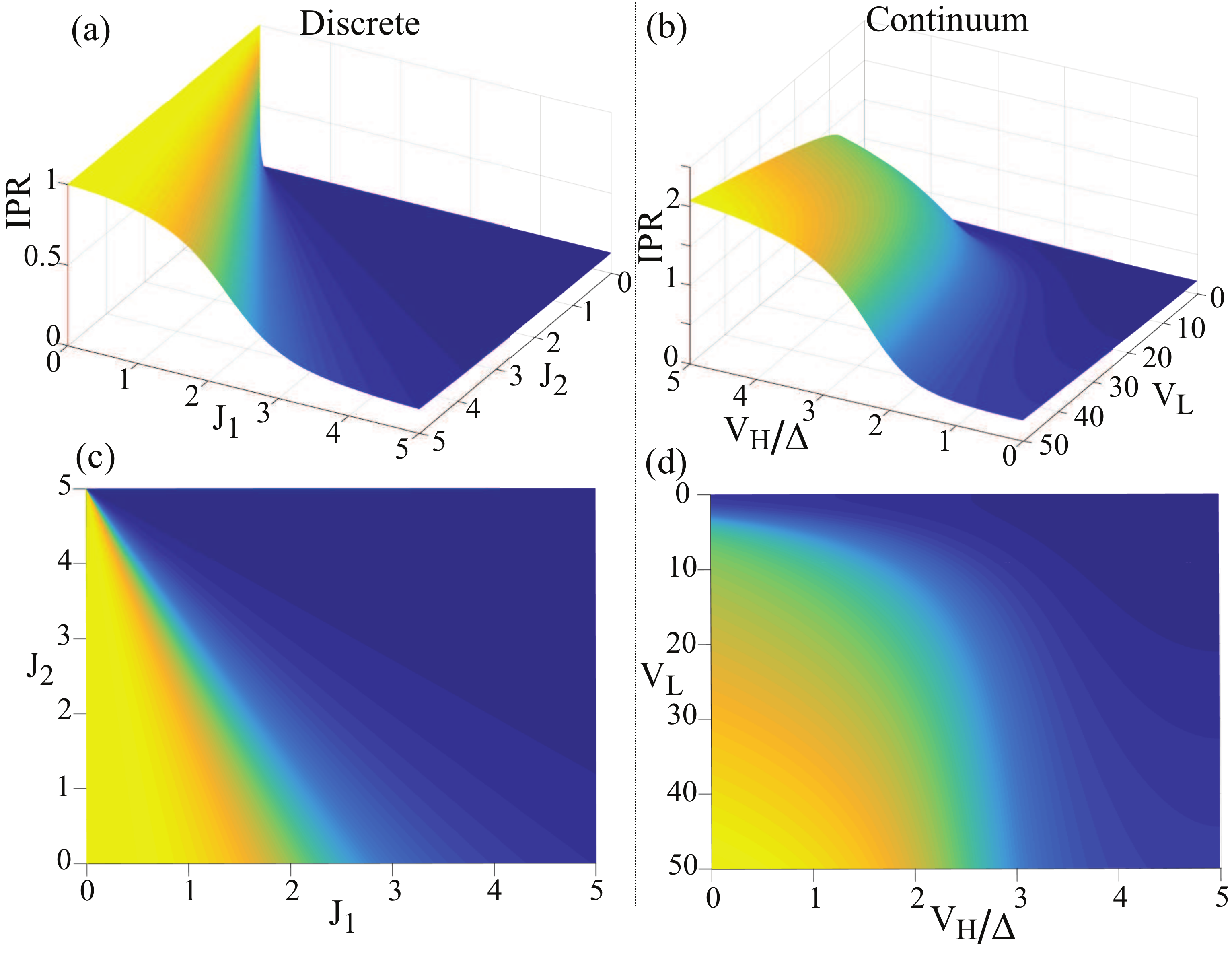}
\caption{The inverse participation ratio (IPR) of the ground state is plotted for $\alpha=5/8$ and $N=8$ lattice sites, for (a) the discrete Hamiltonian, Eq.~(\ref{discrete-Hamiltonian}) as a function of  strengths of the coupling $J_1$ and the modulation $J_2$,  and (b) for the continuum Hamiltonian, Eq.~(\ref{continuum-Hamiltonian}) as a function of the primary lattice amplitude $V_L$ and the continuum equivalent $V_H/\Delta$ of the ratio $J_2/J_1$.  Subplots (c) and (d) show a top view of the same. The localization transition is clearly visible in the change in the IPR, with higher values corresponding to greater localization. }\label{Figure-5-localization}
\end{figure}

Thus, we considered a discrete Hamiltonian of period 8 and choose $\alpha=5/8$,  where the numerator and denominator are belong to the Fibonacci sequence, and computed the inverse participation ratio, given by
\bn IPR=\frac{\sum_n|\phi_n|^4}{(\sum_n|\phi_n|^2)^2}\en
the sum being over the lattice sites of the discrete Hamiltonian. Higher values of the IPR indicate localization and lower values correspond to the extended state. We plot the the IPR for the ground state of the system as a function of $J_1$ and $J_2$ in Fig.~\ref{Figure-5-localization}(a) and (c). There is clearly a localization transition along a line representing the ratio $\lambda=J_2/J_1=2$. We found that the transition becomes sharper as the lattice size is increased. Although not reproduced here, we also computed the IPR averaged over all the states in the band, and it showed a similar localization behavior, albeit a bit more gradual.

We next did the same for our continuum model, where we pick a bichromatic lattice with 8 minima for the primary lattice and $\alpha=5/8$, so that the periods are commensurate, a necessity in a continuum ring configuration. Likewise we compute the IPR, where now the sums are replaced by integrals, and plot it for the ground state in Fig.~\ref{Figure-5-localization}(b) and (d).  Differently from the discrete case, we plot versus the ratio $\lambda_c=V_H/\Delta$ and the primary lattice depth, $V_L$. The localization transition is manifest along the line $\lambda_c=2$.  At very low values of $V_L$, as is to be expected, the localization is lost.

\begin{figure}[t]
\includegraphics[width=\columnwidth]{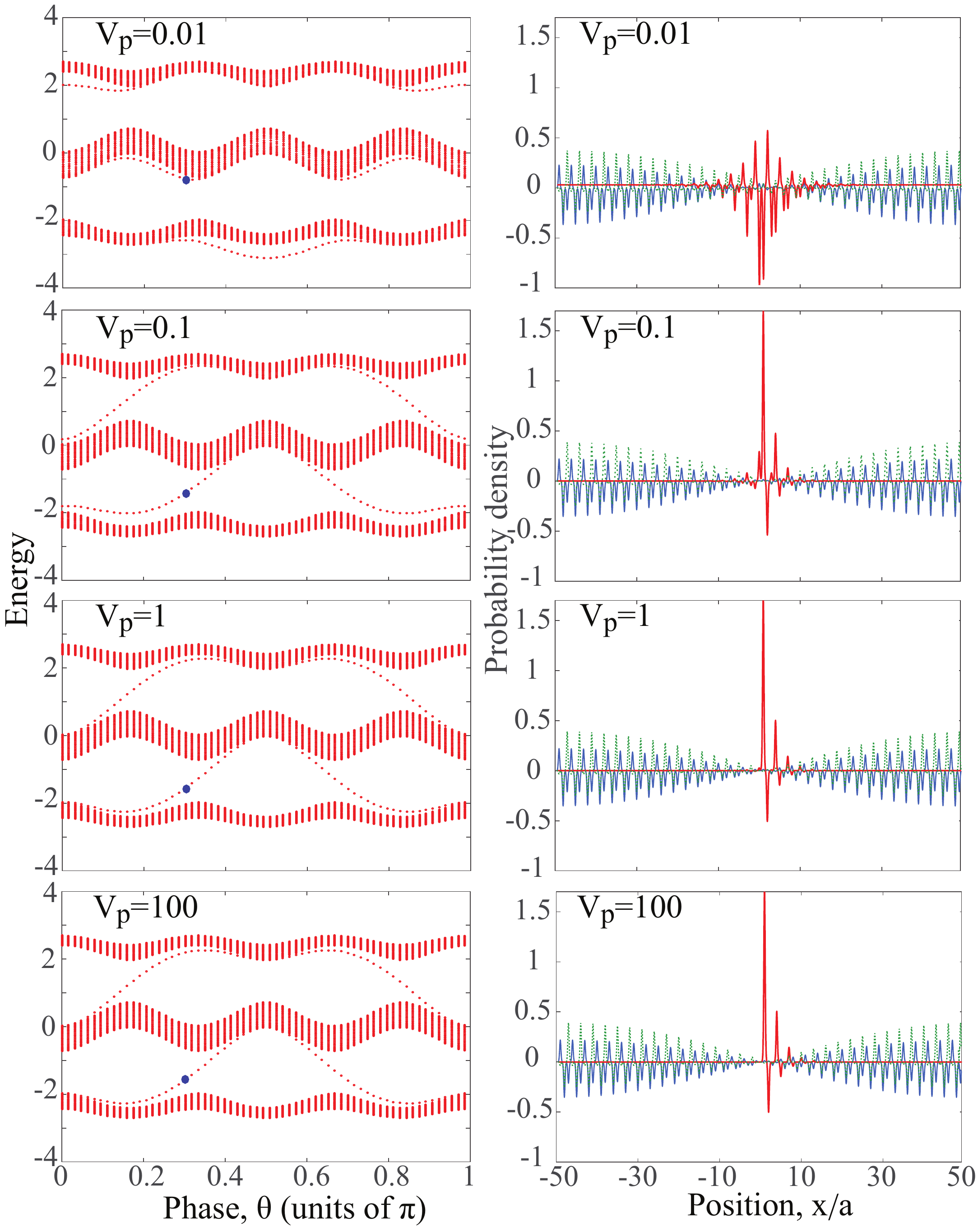}
\caption{The \emph{left} panels show spectra with a Gaussian perturbation, Eq.~(\ref{perturbation}), localized at the bottom of one of 99 wells of the primary lattice in a ring configuration. The large blue dot on each stripe marks the energy corresponding to the localized `edge' state plotted in the corresponding \emph{right} panel. The right panels also show a pair of extended states (one being in dotted line) that have energies immediately adjacent to the edge state, in the bands just above and below. }\label{Figure-6-stripes1}
\end{figure}

\begin{figure}[t]
\includegraphics[width=\columnwidth]{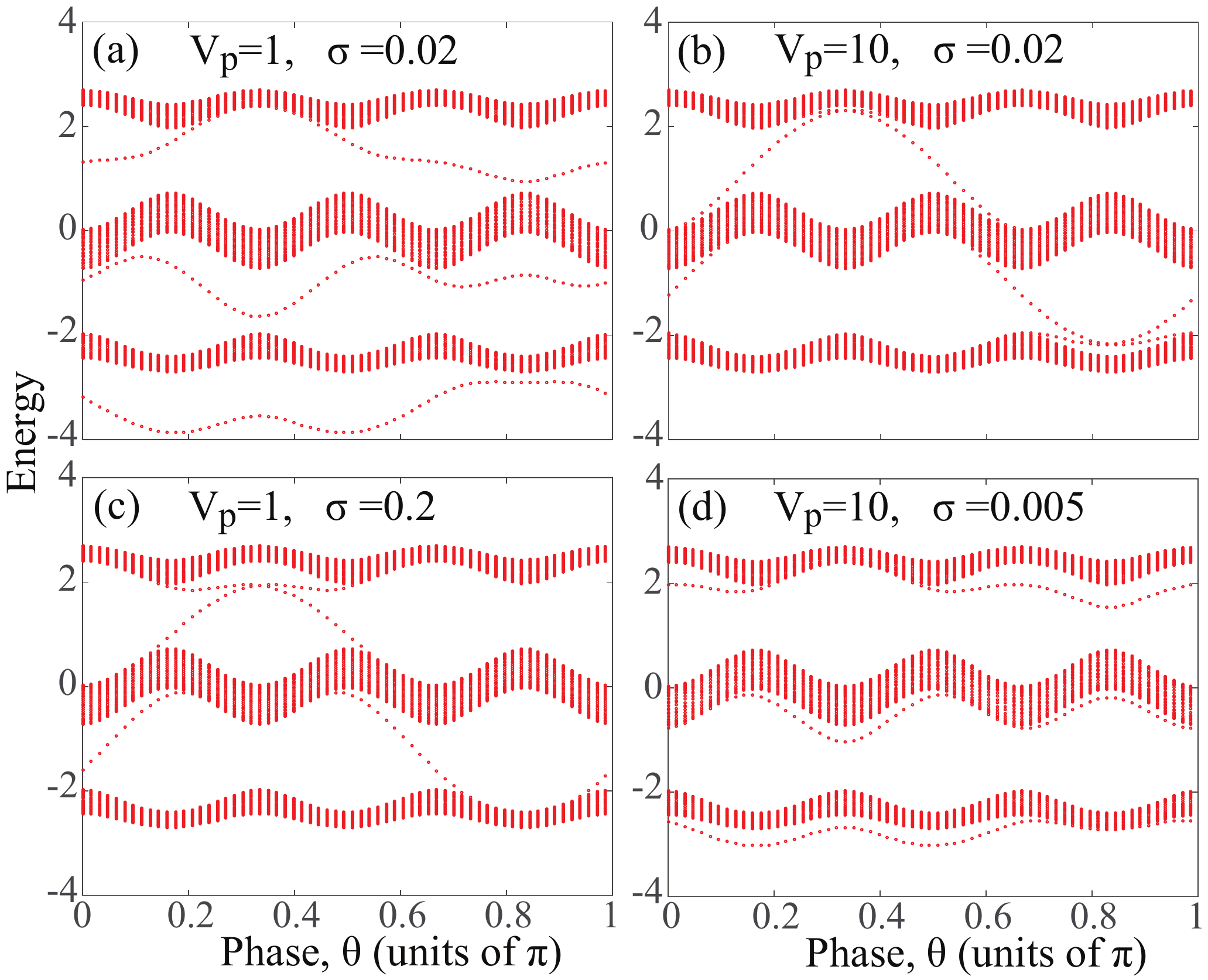}
\caption{The spectrum is plotted with a Gaussian perturbation similarly to Fig.~\ref{Figure-6-stripes1} but now located at one of the crests ($x_0=0.5a$) of the primary lattice. The stripe pattern shows much greater sensitivity to the strength of the perturbation as it is changed from (a) $V_p=1$ to (b) $V_p=10$ with width fixed at $\sigma=0.02$; as well as to the width as it is changed from (a) $\sigma=0.02$ to (c) $\sigma=0.2$. However, increasing the strength while reducing the width simultaneously appears to somewhat mutually cancel the effects as seen by comparing (d) with (a). }\label{Figure-7-stripes2}
\end{figure}

\section{Perturbation and Edge states}

Edge states with energies that exist in the band gaps have been of particular relevance in the physics associated with the Harper model, particularly in the context of the quantum Hall effect \cite{Laughlin-PRB-QHE,TKNN} and in recent years have been intrinsic to exciting developments associated with topological insulators \cite{RMP-Hasan-Kane, RMP-Ryu}.  Counterparts of such edge states in the Aubry-Andr\'e model in an open lattice have been discussed in the literature \cite{Brouwer-stripes,Chen-edge-states} and the localization of the states near the edges have been noted. But, by its very nature the 1D ring configuration corresponds to electrons in an infinite 2D lattice with no edges. However, edges can be mimicked in the lattice simply by `cutting' the ring. Though this could be done in various ways \cite{cut-ring}, here we assume the simple expedient of a localized repulsive perturbation which for large enough strengths would amount to cutting the ring and creating an `edge'. Modeling a barrier on a tightly-focussed blue-detuned laser we use a narrow Gaussian barrier potential,
\bn V_{pert}=V_P\ e^{-(x-x_0)^2/\sigma^2}\label{perturbation}\en
with the width $\sigma$ chosen to be comparable to the lattice period. Doing so introduces eigenvalues in the band gaps exactly as is the case with edge states.  This is demonstrated in Fig.~\ref{Figure-6-stripes1} which shows the appearance of stripes of eigenenergies in the gaps as the phase $\theta$ of the Harper modulation in Eq.~\ref{continuum-Hamiltonian} is varied. Here, we used a perturbation centered at the bottom of a well, specifically $x_0=0$, and of width $\sigma=0.02a$, narrow compared to the lattice period and therefore well-localized within a well.

Just like with edge states, the states corresponding to those eigenvalues in the gap are sharply  localized around the position of the perturbation, as shown in the right panels in Fig.~\ref{Figure-6-stripes1}. This is in stark contrast to the states with energies just above and below in value, which are shown to be completely delocalized in the same plots.

We have used 99 sites with $\alpha=1/3$, which allows the modulation to be commensurate.  Even for moderate perturbation, the spectrum from  the continuum ring model, including the stripe pattern, resembles that for an open 1D lattice for the discrete model in Ref.~\cite{Chen-edge-states}, where the same value of $\alpha$ was used, but with 100 sites. But, if we use 100 sites on a continuum ring, the two potentials in Eq.~(\ref{continuum-Hamiltonian}) become incommensurate, and  other stripe features appear since the mismatch of the lattice around the ring acts like an additional perturbation.

Figure~\ref{Figure-6-stripes1} illustrates a surprising feature: The strength of the perturbation can be very weak compared to the depth of the primary lattice. Here we used $V_L=100$, and even for $V_P/V_L=10^{-4}$, the stripes and localization are already emergent. Furthermore, from $V_P/V_L=10^{-3}$ as the perturbation strength is increased by several orders of magnitude, those features are remarkably invariant, with very little change even quantitatively. The small perturbation strength required makes sense in one way; it is of the same order of magnitude as the Harper modulation used here $V_H=0.024$.  Yet, the strong similarity in behavior with edge states raises questions about the nature of the latter since, with such a weak perturbation, the ring can hardly be considered `cut' and by no means creates the analog of an edge since that would imply the presence of an infinite potential. It appears that certain features like localization and intra-band energies associated with edge states can actually be induced with a very tiny perturbation.

There is another interesting effect that emerges from an additional freedom in the continuum model not present in the discrete model: The position of the perturbation can be varied within the span of of a single period of the primary lattice and specifically it can be positioned at one of its crests so that the perturbation is actually in between two lattice sites, something not literally possible in the discrete model.  We find that the stripe pattern generated for this midway location is much more sensitive to the strength of the perturbation, Fig.~\ref{Figure-7-stripes2}(a) and (b) show that the pattern changes completely when the perturbation strength is increased from $V_P=1$ to $V_P=10$. This is in stark contrast with Fig.~\ref{Figure-6-stripes1} where the perturbation is at the well bottom and the pattern hardly changes over several orders of magnitude variation of $V_P$.

To probe this farther, in Fig.~\ref{Figure-7-stripes2}(c) we kept the strength fixed at $V_P=1$ and instead increased the width $\sigma$ by a factor of 10, and that led to a pattern similar to increasing the strength by a factor of 10. On the other hand, reducing $\sigma$ while increasing $V_P$ seems to compensate for each other, as seen in Fig.~\ref{Figure-7-stripes2}(d) which appears qualitatively similar to Fig.~\ref{Figure-7-stripes2}(a).  It appears that the midway location acts as if in the discrete limit there is a perturbing potential at two adjacent sites, a fact accentuated by strengthening or widening the perturbation.

\begin{figure}
\includegraphics[width=\columnwidth]{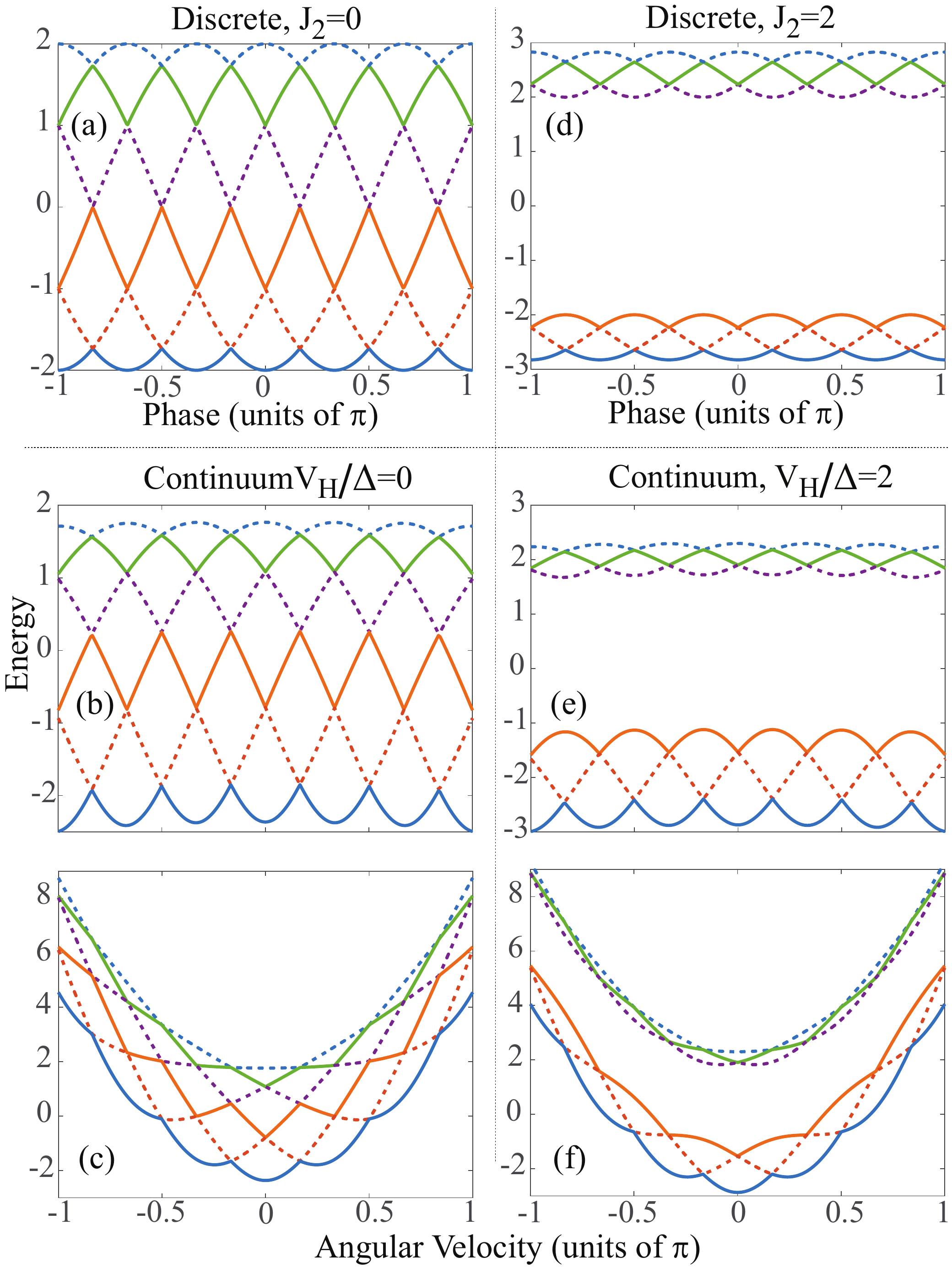}
\caption{The complex coupling $J_1e^{i\vartheta}$ in the Harper Hamiltonian creates oscillations in the spectrum as shown here for the lowest band with N=6 lattice sites and $J_1=1$. The spectrum for the discrete Hamiltonian is shown (a) with no modulation $J_2=0$ and (d) with modulation at the critical strength $J_2=2$ with $\alpha=1/2$.  This behavior is reproduced in the continuum lattice by modifying the momentum operator with a minimal coupling gauge potential term as done in Eqs.~(\ref{minimal-coupling}) and (\ref{rotation}) and plotted here in (b) and (e). The oscillations in the spectrum can be generated by rotating the lattice as seen in (c) and (f), but with a curvature arising from leaving out the square term in Eq.~(\ref{rotation}). The colors and alternating solid and dotted lines differentiate the different energy levels.}\label{Figure-8-rotation}
\end{figure}

\section{Complex Coupling and Rotation}

So far we have neglected the phase $\vartheta$ that can lead to a complex coupling coefficient $J_1e^{\pm i\vartheta}$ in the Harper equation in Eq.~(\ref{discrete-Hamiltonian}).  The phase can originate from one of the wavenumbers $k$ in the 2D problem, with $\vartheta=ka$. Assuming this dependence on the lattice parameter, $a$, in the continuum limit of $a\rightarrow 0$, a Taylor expansion readily establishes the correspondence,
\bn && e^{i\vartheta}\psi(x+a)+e^{-i\vartheta}\psi(x-a)-2\psi(x) \n\\ \simeq && -\frac{a^2}{\hbar^2}\left(\hat{p}-\frac{\hbar\vartheta}{a}\right)^2\psi(x).\label{minimal-coupling}\en
This suggests that in the continuum we simply need to modify the kinetic energy by the minimal coupling that is the standard approach for introducing gauge fields in quantum mechanics. This amounts to the following modification in the continuum Hamiltonian Eq.~(\ref{continuum-Hamiltonian}),
\bn -\frac{\hbar^2}{2m}\frac{d^2}{d x^2}\rightarrow -\frac{\hbar^2}{2m}\frac{d^2}{d x^2}+i\vartheta\frac{\hbar^2}{ma}\frac{d}{d x}+\frac{\hbar^2\vartheta^2}{2ma^2}\label{rotation}\en
The second term can be written as $i\hbar R\Omega d_x$,  with R being the radius of the ring, $\Omega=\hbar\vartheta/(m a R)$ would be the angular velocity of the ring rotating around its symmetry axis through its center. Thus, the effect of this term can be simulated by simply rotating the ring. This is to be expected from the well-known analogy of rotation and magnetic vector potential \cite{Review-Goldman}. The square term has the form of a centripetal contribution, $m R^2\Omega^2$, but has no clear counterpart in the 1D dynamics. For constant angular momentum it would simply provide an energy shift but it has a more significant effect when the angular momentum is varied, as we will see.

In Fig.~\ref{Figure-8-rotation} we show that with the introduction of these extra terms, the effect of the complex coupling can be reproduced faithfully.  In these particular plots, we have further scaled all lengths by $R$. When the Harper modulation is absent, the results for the discrete plot in Fig.~\ref{Figure-8-rotation}(a) and the continuum plot with the two added terms in Fig.~\ref{Figure-8-rotation}(b) are practically identical. The different colors mark the six different energy levels in the first band, due to using $N=6$ cells here, and that is also reflected in the periodicity of the oscillation of the energies as a function of the phase $\vartheta$ and the angular velocity $\Omega$ respectively in plots Fig.~\ref{Figure-8-rotation}(a) and (b). However, when the $\Omega^2$ term is left out in Fig.~\ref{Figure-8-rotation}(c), the spectrum acquires an overall curvature proportional to $\Omega^2$, although other features of the spectrum are preserved. This would be the form of the spectrum if the ring shaped lattice is simply rotated, without any mechanism to simulate the square term.

With the phase and its continuum counterparts in place, if we now include the Harper modulation, the first band breaks into sub-bands, two of them in this case since we use $\alpha=1/2$. The result for the discrete model is shown in Fig.~\ref{Figure-8-rotation}(d) and the continuum model with and without the square terms are shown in Fig.~\ref{Figure-8-rotation}(e) and (f), both of which show qualitatively similar structure.  Apart from the emergence of the sub-bands the behavior is similar to when the modulation is absent, just as discussed above. Notably, there is a  difference in the widths of the two sub-bands in the continuum plots, which arises from the fact that we have used a low value of lattice depth here $V_L=10$, and that difference diminishes at higher lattice depths.

\section{Conclusions and Outlook}

We have examined the continuum version of the Harper model as mapped to a 1D ring-shaped lattice with two commensurate sinusoidal potentials. We have demonstrated that all the salient features of the model can indeed be realized just as well as in the 2D lattice systems, that have had primacy in experiments. By not assuming a tight-binding model, we have identified departures from the idealizations intrinsic in the standard from of the Harper model. Specifically, we observed deviations from bilateral symmetry and general distortions of the signature Hofstadter butterfly spectrum, and found the conditions necessary to approach the idealized picture.

We demonstrated that the localization transition predicted by Aubry and Andr\'e, proven for incommensurate lattice periods, can also be realized on a ring even though the lattice periods are commensurate with rational $\alpha=p/q$ and even when the integers $p,q$ have single digit values. By introducing even a small perturbation, we also found that analogs of localized edge states can be created with much flexibility, possessing the features of the edge states in open systems. Rotating the lattice allows modeling the complex nearest neighbor couplings in the general Harper model, but with a curved spectrum.

Introducing nonlinearity due to atom-atom interaction for bosons in this model will certainly lead to additional features. Some aspects can be surmised from generalization of the discrete 2D model to include a two body interaction term in a Hofstadter-Hubbard model \cite{Iskin-Hofstadter-Hubbard}.  Extending such considerations to a continuum ring system as examined here, can be a fruitful line of future research.

The technology for realizing the physics discussed here already exists. Ring-shaped lattices have been demonstrated with interfering Laguerre-Gaussian (LG) beams that carry opposite angular momentum (OAM). Using beams with two different OAM can create the bichromatic azimuthal lattice structure necessary. Trapping ultracold atoms in LG beams has also been successfully demonstrated in experiments. Therefore, it is primarily a matter of bringing the relevant capabilities together, and we hope this paper can provide some motivation towards that.  Considering how significant the Harper model has been and continues to be in physics, the ability to examine it experimentally in a feasible alternate configuration with some decided advantages, will certainly be a valuable addition to the arsenal of ultracold atomic systems for probing fundamental physics.

\begin{acknowledgments} We gratefully acknowledge discussions with D. Schneble, and the support of the NSF under Grants No. PHY-1313871 and PHY-1707878. \end{acknowledgments}

\vfill

\end{document}